\documentclass[12pt]{article}
\usepackage{amsfonts,amsmath}
 \usepackage{amssymb}

\makeatletter \@addtoreset{equation}{section} \makeatother
 
  \newcommand{\dddag}{{  \begin{picture}(5,5)
\put( 0,0){$\dagger$} \put(1.5,0){$\dagger$}
 \end{picture}}}

\newcommand{\nn}{{\nonumber}}
\newcommand{\rhs}{{\it r.h.s.} }

\newcommand{\KK}{{  \mathcal{K}} }

\newcommand{\p}{{ \partial} }
\newcommand{\bbp}{{ \bar{\partial}} }
\newcommand{\xx}{{\bf x}}
\newcommand{\Z}{{\mathcal{X}}}
\def\dr{\mathrm{d}}
\def\be{\begin{equation}}
\def\ee{\end{equation}}
\def\bea{\begin{eqnarray}}
\def\eea{\end{eqnarray}}
\newcommand{\dgr}{\ddag}
\newcommand{\ff}{\frac}
\newcommand{\gd}{\delta}
\newcommand{\gl}{\lambda}
\newcommand{\pl}{\dot{\lambda}}
\newcommand{\bmu }{\bar{\mu}}
\newcommand{\gk}{\kappa}

\newcommand{\gvep}{\varepsilon}
\newcommand{\go}{\omega}

  \newcommand{\half}{{\ff{1}{2}}}
 \newcommand{\gs}{\sigma}

\newcommand{\bz}{\bar z}

\newcommand{\iz}{  {\mathrm{z}}}
\newcommand{\iy}{  {\mathrm{y}}}
\newcommand{\Y}{{\textrm {Y}}}
\newcommand{\biz}{ \bar{\mathrm{z}}}
\newcommand{\biy}{ \bar{\mathrm{y}}}

\newcommand{\by}{{\bar{y}}}

\newcommand{\bk}{{\bar{k}}}

\newcommand{\bgx}{{\bar{\xi}}}

\newcommand{\q}{\,,\qquad}

 \newcommand{\eq}{\eqref}

\newcommand{\bee}{\begin{eqnarray}}
\newcommand{\beee}{\begin{array}}
\newcommand{\eee}{\end{eqnarray}}
\newcommand{\eeee}{\end{array}}

\newcommand{\gx}{\xi}

\newcommand{\ga}{\alpha}
\newcommand{\pa}{{\dot{\ga}}}
\newcommand{\pb}{{\dot{\gb}}}

\newcommand{\gb}{\beta}
\newcommand{\gga}{\gamma}
\newcommand{\bga}{\bar{\gamma}}
\newcommand{\ls}{\!\!\!\!\!\!}
\newcommand{\bkap}{\overline{{\kappa}}}
\newcommand{\ie}{{\it i.e.,} }

\begin{document}
\begin{flushright}
\vspace{1mm}
FIAN/TD/ №17-26\\
\end{flushright}

\vspace{1cm}

\vspace{0.5cm}
\begin{center}
{\large\bf  Nonlinear Higher-Spin Theory in $O(4,1)$ de Sitter Space}

 \vglue 0.6  true cm

\vskip0.5cm

 O.A. Gelfond$^1$ and M.A.~Vasiliev$^{1,2}$
 \vglue 0.3  true cm

\vspace{1 cm}

  \vspace{0.5cm}

 \textsc{$^{1}$} \textit{I.E. Tamm Department of Theoretical Physics,
Lebedev Physical Institute,}\\
 \textit{ Leninsky prospect 53, 119991, Moscow, Russia}\\

\par\end{center}

\begin{center}
\textsc{$^{2}$}\textit{Moscow Institute of Physics and Technology,}\\
 \textit{ Institutsky lane 9, 141700, Dolgoprudny, Moscow region,
Russia}\\

\par\end{center}

 \vspace{1 cm}

{\it $\phantom{MMMMMMMMMMMMMMMMMM}$ To the memory of  Igor Tyutin
$\phantom{MMMMMMMMMMMMMMMMMMMMMMMMMMMMMM}$}
\vspace{1 cm}

\begin{abstract}
Four versions of the  higher-spin gauge theory compatible
with $O(4,1)$ de Sitter
geometry in presence of fermions are proposed. Two of them exhibit the structure of
$\mathbb{Z}_2\times \mathbb{Z}_2$ Lie superalgebra proposed originally by one of the authors for the description of de Sitter supergravity \cite{Vasiliev:1985vtv}. Two of the proposed models are unitary in the bosonic sectors in analogy with the recent proposal by
Lang and Neiman \cite{Lang:2026inn} and  admit a pseudo involution in
presence of fermions suggesting possible (anti)unitarity
 associated either with the time reversal in $dS_4$ or with the space-time
 doubling.
 \end{abstract}

 \newpage
 \tableofcontents

\newpage

\section{Introduction}

Usually, higher-spin (HS) gauge theory is formulated in the $O(d-1,2)$
invariant anti-de Sitter ($AdS_d$) space, where  it consistently describes
HS gauge fields of both integer and half-integer spins. (For the reviews of HS theories  see e.g. \cite{Vasiliev:1999ba,Bekaert:2004qos}.) In HS theory,
the cosmological constant $\Lambda$ is expressed via parameter $\lambda$,
\be
\Lambda =-\lambda^2\,,
\ee
so that $\lambda$ is real in $AdS_d$ but pure imaginary in the de Sitter ($dS_d$)
case. In the bosonic theory
 only even powers of $\lambda$ contribute. However, in presence of
fermions,  the field equations  involve odd powers of
$\lambda$ and the theory becomes ill-defined due to imaginary mass-like parameters resulting in a complex action.
This phenomenon is not specific for HS theory, taking place
already in supergravity, that is well-defined in $AdS$ but not in $dS$.

Nowadays, the interest to $dS$ geometry
(see \cite{Mukhanov:1981xt}-\cite{Han:2026hrd} for an incomplete list of references)
is mostly because, as it became clear relatively
recently \cite{Sahni:1999gb}, it properly models the dark energy phenomenon of the present stage of Universe evolution.
Since $dS$ Universe is not stationary
(expanding), some of the principles of the standard QFT may be relaxed including
the absence of ghosts and boundness of the energy especially at the quantum gravity regime in which HS gauge fields may play a key role.
 Indeed, it is well known that
algebra $o(d,1)$ admits no lowest weight (\ie finite spin and bounded energy) unitary representations \cite{HarCha}.

 A related fact is that
the $AdS_4$ symmetry algebra $sp(4)\sim o(3,2)$ admits a supersymmetric extension  $osp(1,4)$ while $o(4,1)$ does not.
Nevertheless some versions of SUGRA in $dS_4$ were proposed in \cite{LukNow2,PilchAtall,Vasiliev:1985vtv}.
 All of these models do not respect unitarity.\footnote{In this paper we focus on the models with unbroken SUSY. There are models of supergravity with
broken supersymmetry around $dS$
vacuum (see e.g. \cite{Volkov:1973ix,Bergshoeff:2015tra,Ferrara:2016ajl}). So far such models are unavailable in the HS theory since spontaneous breaking of HS symmetries has not been worked out.
(See however \cite{Didenko:2023txr,Didenko:2025xca}.)}

The construction of \cite{Vasiliev:1985vtv} was characterized by two
related features. Firstly, it involved so-called Klein operators
that play a major role in the HS theory \cite{Vasiliev:1988sa,Vasiliev:1992av}. Secondly,
the model of $o(4,1)$ SUGRA of \cite{Vasiliev:1985vtv} was based on the
$\mathbb{Z}_2\times \mathbb{Z}_2$ generalized superalgebras introduced in \cite{rw1,rw2}
and extensively studied more recently in \cite{tol}-\cite{aaca}.

Another reason for increased  interest to HS theory in $dS_4$ is that it provides
a useful model for studying a $dS_4$/CFT holography counterpart of the $AdS/CFT$ holography \cite{Maldacena:1997re}-\cite{Klebanov:2002ja}
 as discussed by many authors \cite{{Strominger:2001pn},{Spradlin:2001pw},{Anninos:2011ui}, {Chang:2013afa},{Arias:2019pzy},{Arias:2019zug},{David:2020ptn},
 {Arenas-Henriquez:2022pyh},{Dey:2024zjx}}.
  So far, bosonic HS theories were mostly discussed from that perspective
  because of the lack of their supersymmetrization in $dS_4$ with the exception of the paper by Sezgin and Sundell
\cite{Sezgin:2012ag} where a $dS_4$ version of the $AdS_4$ supersymmetric HS algebra of \cite{Konstein:1989ij} was defined in a way analogous to that of
the approach of \cite{LukNow2,PilchAtall} to SUGRA.

The aim of this paper is to propose nonlinear models of
$dS_4$ HS theory that differ from the $AdS_4$ model of \cite{Vasiliev:1992av} by
hermiticity conditions involving the Klein operator. The idea of the construction is inspired by that of \cite{Vasiliev:1985vtv}. However, its realisation  is even simpler in the HS theory  because of the doubled number of fermions due to the presence of the Klein operators.

The nonlinear equations for $4d$ massless fields of all spins  of
 \cite{Vasiliev:1992av} with standard hermiticity conditions describe
a HS theory with  $AdS_4$ as the most symmetric vacuum.
We show that the same  equations with appropriately modified Hermiticity conditions describe HS theory with $dS_4$ vacuum.
In fact, we propose several types of $dS_4$ HS theories: some analogous to the constructions of \cite{LukNow2,PilchAtall,Sezgin:2012ag} and others   to \cite{Vasiliev:1985vtv}. In each case we consider two types of reality
conditions, one leading to a non-unitary theory and another one leading to a unitary theory in the bosonic sector and endowed with the pseudo involutive  conjugation that squares to $-1$ on fermions,
$
(a^\dag)^\dag =(-1)^{\pi(a)} a\,.
$
Note that this type of conjugations allows one to escape a contradiction with the  representation theory of $O(4,1)$ \cite{HarCha} not allowing positive bounded energy unitary representations, the existence of which would be a consequence of positive definiteness of $\{Q_\ga\,,\bar Q_\pa\}$.
Also, it agrees with the existence of the pseudo involution on
spinors on the $SU(2)/U(1)$ boundary of $dS_4$.

 The latter type of reality conditions is analogous to the one proposed  in the interesting recent paper \cite{Lang:2026inn} for a unitary version of the $dS_4$ HS theory with bosonic fields of all spins.
Its extension to the full HS theory with fermions is achieved via specific
generalization of the reality conditions with the aid of the Klein operators present in this paper.

Let us stress that the full nonlinear HS gauge theory with fermions needs the exterior
 Klein operators that anticommute with the differentials $\theta^\ga = dz^\ga$ or $\bar \theta^\pa =d\bar z^\pa$ of \cite{Vasiliev:1992av} (for detail see next section). These can be avoided in the nonlinear bosonic subsector or at the linearized level but not in the full nonlinear theory with fermions.
For that reason our construction of the nonlinear HS field equations with fermions differs from that of \cite{Sezgin:2012ag}.

The layout of the rest of the paper is as follows.
 In   Section \ref{Preliminaries}, the unfolded form of $4d$ free and nonlinear
 HS  equations in $AdS_4$ proposed in
\cite{Vasiliev:1988sa,Vasiliev:1992av} is recalled.
General properties and particular examples of discrete maps, that act on
solutions of the HS equations, are briefly recalled in Section \ref{inv}
with the emphasize on reductions and Hermiticity conditions.
The Hermiticity conditions of the $AdS_4$ HS
theory are recalled in Section \ref{AdS4 Hermiticity}.
In Section \ref{I},  a  modification of the Hermiticity conditions to the $dS_4$ HS theory with fermions referred to as type-I $dS_4$ HS theory is proposed.  In  Section \ref{Z2Z2}
it is shown that the structure of $\mathbb{Z}_2\times \mathbb{Z}_2$ Lie superalgebra introduced in
\cite{Vasiliev:1985vtv} for the description of $dS_4$ SUGRA admits a natural
realization in  the nonlinear HS system of \cite{Vasiliev:1992av}. This results in a type-II $dS_4$ HS theory. Unitarity of the proposed models is analysed in Section \ref{uni}, where pseudo unitary HS theories in $dS_4$ are identified.
Section \ref{conc} contains summary of the obtained results along with the brief discussion of further research directions including HS $dS_4/CFT$  holography.
\section{Conventional HS theory in $AdS_4$}
\label{Preliminaries}
 The formulation of \cite{Vasiliev:1992av} of the HS theory
 in $AdS_4$ uses the language of spinors.
Its relation to the conventional setup in
terms of space-time derivatives is via unfolded equations which we briefly
recall now. These are formulated in terms of the
one-form $  \omega (Y;K| x)= dx^n\omega_n (Y;K| x)$ for the gauge fields of all spins
and zero-form $ C(Y;K| x)  $ for HS field strengths and matter fields \cite{Vasiliev:1988sa}, $Y^A=(y^\ga,\by^\pa)$.\footnote{$A=1,\ldots 4$ is a Majorana spinor
index
while $\ga = 1,2$ and $\pa =1,2$ are two-component ones raised and lowered
by $\varepsilon_{\ga\gb}=-\varepsilon_{\gb\ga}$, $\varepsilon_{12}=1$: $A^\ga =\gvep^{\ga\gb} A_\gb$,
$A_\ga = A^\gb\gvep_{\gb\ga}$ and analogously for dotted indices.}
Klein operators $K=(k,\bar k)$ are defined to induce chirality automorphisms
via relations
\be
\label{kk}
k y^\ga = -y^\ga k\,,\quad
k \bar y^\pa = \bar y^\pa k\,,\quad
\bar k y^\ga = y^\ga \bar k\,,\quad
\bar k \bar y^\pa = -\bar y^\pa \bar k\q kk=\bar k\bar k = 1\,,\quad
k\bar k = \bar k k\,.
\ee
To describe massless fields, the one-form $  \omega (Y;K| x)$ and zero-form $C (Y;K| x)$
have to be, respectively, even and odd in $k,\bar k$,
\be\label{Csumkbark}
C(Y;K|x)= C^{1,0}(Y|x) k + C^{0,1}(Y|x) \bar k\q \go(Y;K|x)= \go^{0,0}(Y|x)  + \go^{1,1}(Y|x) k\bar k\,.
\ee
As a result, massless fields are doubled in the (unreduced) HS theory. This doubling
will be important in the construction of this paper.

Unfolded  field equations for free massless fields of all spins in $AdS_4$ read as  \cite{Vasiliev:1988sa}
\bee
\label{CON1}
    \ls\ls\ls R_1(Y;K| x)&\!=\!&
\ff{i}{4} \!\Big (\! \eta \overline{H}^{\pa\pb}\bbp_\pa\bbp_\pb \
{ C}(0,\overline{y};K| x) k +\bar \eta H^{\ga\gb}\! \p_\ga\p_\gb\!
{C}(y,0;K| x) \bar k\!\Big ) , \quad\\\label{CON2}
\, \qquad\tilde{D}C (Y;K| x) &\!=\!& 0\,, \eee
where  $\p_\gb:=\ff{\partial}{\partial  {y}^\gb}$, $\bbp_\pb:=\ff{\partial}{\partial \bar{y}^\pb}$, $\eta$
is a free complex parameter and
\bee
\label{RRR}
&&\ls\ls\ls R_1 (Y;K| x) :=D^{ad}\omega (Y;K |x) :=
D^L \go  (Y;K| x) +
\lambda h^{\ga\pb}\Big (y_\ga \bbp_\pb
+   \p_\ga \bar{y}_\pb\Big )
\omega  (Y;K | x) \,,
\\\label{tw}
&&\ls\ls\ls \tilde D C(Y;K |x) :=D^{tw}C (Y;K |x) :=
D^L C (Y;K |x) -{i}\lambda h^{\ga\pb}
\Big (y_\ga \bar{y}_\pb -\p_\ga\bbp_\pb\Big ) C (Y;K |x)\,,
\\\label{dlor}
&&\ls\ls\ls D^L f (Y;K|x) :=
\dr_x f (Y;K|x) +
\Big (\go_L^{\ga\gb}y_\ga \p_\gb +
\overline{\go}_L^{\pa\pb}\bar{y}_\pa \bbp_\pb \Big )f (Y;K|x),\quad
\dr_x: =dx^n\ff{\p}{\p x^n}
 .\,\,\,
\eee
Here $D^{ad}$ (\ref{RRR}) and $\tilde D$ (\ref{tw}) are the adjoint and twisted adjoint covariant derivatives.

Background $AdS_4$ space of radius $\lambda^{-1}=\rho$
is described by a flat $sp(4)$
connection $w=(w_{\alpha \gb},\overline{w}_{\pa\pb},h_{\ga\pb})$
containing Lorentz connection
$w_{\alpha \gb} $, $\overline{w}_{\pa\pb}$ and
vierbein  $h_{\ga\pb}$, that obey
\bee
\label{nR}
\dr_x w_{\alpha \gb} +w_{\alpha}{}_\gamma
 w_{\gb}{}^{\gamma} -\lambda^2\, H_{\alpha \gb}=0\,,\quad
\dr_x\overline{w}_{{\pa}
{\pb}} +\overline{w}_{{\pa}}{}_{\dot{\gamma}}
 \overline{w}_{{\pb}}{}^{ \dot{\gga}} -\lambda^2\,
 \overline H_{{\pa\pb}}=0\,,\quad
\\ \label{tors}\dr_x h_{\alpha{\pb}} +w_{\alpha}{}_\gamma
h^{\gamma}{}_{\pb} +\overline{w}_{{\pb}}{}_{\dot{\delta}}
 h_{\alpha}{}^{\dot{\delta}}=0\,.\quad
\eee
Here $H^{\ga\gb} := h^{\ga}{}^\pa  h^\gb{}_\pa$ and $\overline{H}^{\pa\pb} :=
h^{\ga}{}^\pa h_{\ga}{}^{\pb}$ are the frame two-forms.\footnote{The wedge symbol is  omitted in the sequel of this paper;
 all products of differential forms are exterior.}

Note that the $sp(4)$ subalgebra of the HS algebra, spanned by the
the bilinears of $y_\ga$ and $\bar y_\pa$ belongs to a larger $osp(2,4)$
algebra with the supergenerators $Q_{iA}$ and $u(1)$ generator $T$,
\be
\label{N2}
Q_{1A}:= Y_A\q Q_{2A} := i\KK Y_A\q T := i\KK\q
 \KK=k\bk\,.
\ee
To take the flat limit it is useful to perform   rescalings   \cite{Gelfond:2010pm}
within the decomposition
\be
\label{dec}
A(y,\overline{y}\mid x) =
A_+(y,\overline{y}\mid x) + A_-(y,\overline{y}\mid x)
+A_0(y,\overline{y}\mid x)\,,
\ee where
 \be  A_{+(-)}(y,\overline{y} ) = \sum_{n> m(n<m)}  A_{+(-)}{}_{\ga(n)\pa(m)} y^{\ga(n) }\by^{ \pa(m)} \q
   A_0 (y,\by )
= \sum_{n=m} A_0 {}_{\ga(n)\pa(m)} y^{\ga(n) }\by^{ \pa(m)} \,.
\ee
The rescalings  in the adjoint   and twisted
adjoint modules are  different:
\bee
\tilde{A}^{ad}(y,\overline{y}\mid x)=&&\ls
A_+(|\lambda|^{\half} y,|\lambda|^{-\half}\overline{y}\mid x) +
 A_-(|\lambda|^{-\half}y,|\lambda|^{\half} \overline{y}\mid x)
+A_0(y,\overline{y}\mid x)\,,\\  \label{resc}
\tilde{\tilde{A}}^{tw}(y,\overline{y}\mid x)=&&\ls
A(|\lambda|^{\half}y,|\lambda|^{\half}\overline{y}\mid x)\,.
 \eee
For the rescaled fields,
the adjoint and twisted adjoint covariant derivatives (\ref{RRR}),
(\ref{tw}) give
\be \label{rescad}
D_{rsc}^{ad}\tilde{A} := D^L \tilde{A}  +
  h^{\ga\pb} \Big(  y_\ga \bbp_\pb   \tilde{A}_{-}
+\p_\ga \bar{y}_\pb\tilde{A}_{+}+
\gl^2 \big (y_\ga \bbp_\pb  \tilde{A}_{+}
+\p_\ga \bar{y}_\pb\tilde{A}_{-}\big )+\gl \big (y_\ga \bbp_\pb
+  \p_\ga \bar{y}_\pb\big )\tilde{A}_{0}\Big )
 \,,
 \ee
 \be
\tilde D_{rsc}^{tw} \tilde{\tilde{A}} :=
D^L \tilde{\tilde{A}} -{i} h^{\ga\pb}
\big (\gl^2 y_\ga \bar{y}_\pb -\p_\ga \bbp_\pb  \big ) \tilde{\tilde{A}}\,.
\ee
The flat limits $\lambda\to 0$ of  $D_{rsc}^{ad} $ and $D_{rsc}^{tw}$ are well defined.
 For instance, \be
\label{fltw}
D_{fl}^{tw} \tilde{\tilde{A}}(y,\bar{y} \mid x) =
D^L \tilde{\tilde{A}}(y,\bar{y} \mid x) +i h^{\ga\pb}\p_\ga \bbp_\pb
\tilde{\tilde{A}}(y,\bar{y} \mid x)\,.
\ee
The resulting field equations
describe free HS fields in Minkowski space.   (For  detail see  \cite{Gelfond:2010pm}).
Note that, although  the contraction $\lambda\to 0$
 is consistent with the free
HS field equations,  negative powers
of $\lambda$ survive in the full nonlinear equations
 upon  rescaling (\ref{resc}), making
the Minkowski background  unreachable in the
nonlinear HS gauge theories of \cite{Fradkin:1986qy,Vasiliev:1992av,Vasiliev:2003ev}.


The problem is to find a consistent with $\dr^2=0$ and, hence, gauge invariant (see, e.g., \cite{Bekaert:2004qos}) nonlinear deformation of the equations
(\ref{CON1}), (\ref{CON2}). This  is solved by the following nonlinear system
of \cite{Vasiliev:1992av}
\be\label{WW}
\mathrm{d}_{x}W+W*W=0\q
\mathrm{d}_{x}S+W*S +S*W=0\,,
\ee
\be\label{SS}
S*S= i (\theta^A \theta_A + \eta B* \gamma +
 \bar \eta B* \bar\gamma ) \,,
\ee
\be
\mathrm{d}_{x}B+W*B -W*B=0\q
\label{SB}
S*B=B*S\,,
\ee
where
\be\label{gamma=} \gamma=\theta^\ga \theta_\ga  \kappa k=2k\gk\gd^{2}(\theta)\,,\qquad
\bar\gamma=\bar \theta^\pa \bar \theta_\pa \bkap \bar k=2 \bkap \bar k\gd^{2}(\bar\theta)\,,\qquad
,\ee
 $\theta^A:=dZ^A$
are anticommuting $Z-$differentials, $\theta^A \theta^B=-\theta^B
\theta^A$. Here
    \textbf{$B(Z;Y;K|x)$ }is a zero-form, while
 $W=dx^n W_n (Z;Y;K|x)$ and
$
S=\theta^A S_A (Z;Y;K|x) \,
$
are one-forms.

The  associative star product $*$ acts
 on functions of  spinor variables $Y^A$ and $Z^A$
\be
\label{star2}
(f*g)(Z;Y)=
\int \ff{d^{4} U\,d^{4}V}{(2\pi)^{4}}  \exp{[iU^A V^B C_{AB}]}\, f(Z+U;Y+U)
g(Z-V;Y+V) \,.
\ee
It is normalized so that $f*1 = 1*f =f\,$ and
 provides a
realization of the Weyl algebra,
\be \label{WeylAdS}
[Y_A,Y_B]_*=-[Z_A,Z_B ]_*=2iC_{AB}\,,\qquad
[Y_A,Z_B]_*=0\q [a,b]_*:=a*b-b*a\,.
\ee

The  Klein operators satisfy relations analogous to (\ref{kk}) with
 $y^\ga\to w^\ga= (y^\ga, z^\ga, \theta^\ga )$, $\bar y^\pa\to\bar w^\pa =
(\bar y^\pa, \bar z^\pa, \bar \theta^\pa )$, that extend the star product to the Klein operators.
Decomposing master fields  with respect to the  parity in the Klein operators,
 $A^\pm(Z;Y;K|x)=\pm A^\pm(Z;Y;-K|x)$, ($K:=(k,\bar k)$, $-K:=(-k,-\bar k) $)
 HS gauge fields are
 $W^+,S^+$ and $B^-$ while  $W^-$, $S^-$ and $B^+$
 describe an infinite tower of topological fields
 with every $AdS_4$ irreducible field describing at
most a finite number of DoF. (For more detail see
\cite{Vasiliev:1992av, Vasiliev:1999ba}).

The left and right inner Klein operators
\be
\label{kk4}
\kappa :=\exp i z_\ga y^\ga\,,\qquad
 \bar{\kappa}:=\exp i \bar{z}_\pa \bar{y}^\pa\,,
\ee
 which enter Eq.~\eq{gamma=}, change a sign of
 undotted and dotted spinors, respectively,
\be
\label{[uf]}
\kappa *f(z,\bar{z};y,\bar{y})=f(-z,\bar{z};-y,\bar{y})*\kappa\,,\quad
\bar{\kappa} *f(z,\bar{z};y,\bar{y})=f(z,-\bar{z};y,-\bar{y})*\bar{\kappa}\,,
\ee
\be
\kappa *\kappa =\bar{\kappa} *\bar{\kappa} =1\q \kappa *\bar{\kappa} = \bar{\kappa}*\kappa\,
\ee
but commute with the differentials $\theta^A$.

There are two types of central elements on the \rhs of
\eqref{SS}. One is $\theta^ A\theta_A$ which obviously
commutes with anything. The other ones are $\gga$ and $\bar\gga$
\eq{gamma=},
which commute with everything including $\theta^{\ga}$, $\bar\theta^{\pa}$ thanks to the Grassmann $\delta$-function, $\theta^{\ga}\gd^{2}(\theta)=0$.
That $\gga$ and $\bar\gga$ are central is crucial for the consistency of the whole system. The same time, the presence of the Klein operators in the nonlinear HS system makes it nontrivial, \ie inequivalent to a
system of zero-curvature conditions. This property is inherited by the $dS_4$ HS systems of this paper but seems to be lacking in the $dS_4$ suspersymmetric system of \cite{Sezgin:2012ag}.

\section{Involutions, conjugations and (anti)automorphisms}
\label{inv}

The nonlinear HS equations possess discrete operations identified originally
in \cite{Vasiliev:1986qx}. These are
either linear, that do not conjugate complex numbers  (automorphisms $\tau$ and antiautomorphisms $\rho$),  or antilinear,
that conjugate complex numbers (conjugations $\gs$ and involutions $\dagger$).
Automorphisms and conjugations preserve the order of product factors,
\be
\tau (a* b)= \tau(a) *\tau(b)\q \gs (a* b)= \gs(a) *\gs(b)\,,
\ee
while antiautomorphisms and (pseudo)involutions reverse it,
 \be
 \label{rev}
\rho (a* b)= \rho(b) *\rho(a)\q  (a* b)^\dag= b^\dag *a^\dag\,.
\ee
Involutions and conjugations are in addition  demanded to  square to identity.
Pseudoinvolution $\dagger$ is involutive on even elements (bosons) but squares to $-1$ on odd ones (fermions),
\be\label{pinv}
((a)^\dag)^\dag = \tau_f (a)\q  \tau_f(a):= (-1)^{\pi_{a}} a\,.
\ee

If a system of unfolded equations on the fields $\phi$ is invariant under the action of an
automorphism $\tau$ or conjugation $\sigma$
 the conditions
\be
\tau (\phi) = \phi\quad
\mbox{
or}\quad
\sigma (\phi) = \phi\,
\ee
single out, respectively, a consistent truncation or a real form of the complex system in question. To be imposed
simultaneously $\tau$ and $\gs$ have to commute,
$
\tau\circ \gs = \gs\circ \tau\,.
$

The composition of any of the above operations with an automorphism gives an
operation of the same type (possibly not involutive, however). For instance,
\be
\label{dag'}
\dag':\quad a^{\dag'} := \tau (a^\dag)
\ee
defines a new (pseudo)involution.

 HS equations (\ref{WW})-(\ref{SB}) make sense with the fields
 valued in any associative algebra (e.g. $Mat_n$) \cite{Vasiliev:1988sa} that implies the $A_\infty$ structure \cite{stash}. The HS gauge theory admits a variety of reductions associated with its automorphisms. There are two types of automorphisms of the HS theory, resulting from either automorphisms or antiautomorphisms of the original  $A_\infty$ setup. For instance, the boson-fermion  parity automorphism $\tau_f$ is of the first type allowing to consider purely bosonic
reduction while the reduction to the minimal HS model with only
even spins is of the second type being based on the algebra $ho(1,0|4)$ in notations of \cite{Konstein:1989ij,Vasiliev:1999ba}.

For a system based on an associative algebra $A$ like HS equations  (\ref{WW})-(\ref{SB}) one can use antiautomorpisms $\rho$ of $A$ to generate automorphisms $\tau_\rho$ of the Lie superalgebra $Lie(A)$ associated  with $A$ via graded commutators \cite{Vasiliev:1986qx},
\be\label{taurho}
\tau_\rho (\phi):=  i^{\pi_\phi} (-1)^p \rho(\phi)\,,
\ee
where $\phi$ is a $p$-form of the Grassmann grading $\pi_\phi =0,1$. For instance this is the case for the system (\ref{WW})-(\ref{SB}) with one-forms $W$, $S$
and zero-form $B$.
The basis antiautomorphism of the star product (\ref{star2}) with HS fields carrying inner matrix indices is
\begin{equation}
\label{rho}\rho (P)_k{}^l(Z,Y) = (i)^{\pi (P_k{}^l )}\,
\eta^{l u} P_u{}^v(-iZ,i Y )\eta^{-1}{}_{vk}\,,
\end{equation}
where $P_u{}^v(Z,Y)$ are $Z,Y$-dependent matrices of  boson-fermion
parity $\pi(P_k{}^l )=0$~or~1, while $\eta_{vk}$ is a nondegenerate bilinear
form. The  definite symmetry bilinear forms induce reduced HS systems of the types $ho(n,m|4)$ for $\eta_{vk}=\eta_{kv}$ and $husp(n,m|4)$ for $\eta_{vk}=-\eta_{kv}$ \cite{Konstein:1989ij,Vasiliev:1999ba}. Extension of $\rho$ to the Klein operators is via the antiautomorphism property (\ref{rev}).

Note that a system with the pseudo involution can be extended to a doubled system with the involution as follows. Consider fields of the form $\Phi_1 \Pi_1 +\Phi_2 \Pi_2$ with two projectors $\Pi_{1,2}$, $\Pi_i \Pi_j =\delta_{ij} \Pi_j$ and the automorphism
$
\tau_{12}(\Phi_{1,2})=\Phi_{2,1}\,.
$
Let $\dagger$ be a pseudo involution (\ref{pinv}) of the original system.
Then $\dagger'$ defined by
\be
 (\Phi_1)^{\dagger'} = \tau_{12}(\Phi_1^\dagger)\q
 (\Phi_2)^{\dagger'} = \tau_{f}(\tau_{12}(\Phi_2^\dagger))\,
\ee
with the boson-fermion automorphism $\tau_f$ forms an involution of the doubled one.

\section{$AdS_4$ Hermiticity conditions}
\label{AdS4 Hermiticity}  The Hermiticity conditions on the fields  $\go(Y)$ and $C(Y)$ of the $AdS_4$ HS theory are induced
by the involution $\dagger$  acting on functions
\be
f(y,\bar y) = \sum_{n,m} f^{\ga_1,...\ga_n;\pa_1,...\pa_m}y_{\ga_1}\cdots y_{\ga_n}\by_{\pa_1}\cdots \by_{\pa_m}
\ee
 as
\be
\label{invads}
(f(y,\bar y))^\dagger  = \bar f (\bar y, y)\q
\bar f(\bar y, y) = \sum_{n,m} \bar f^{\pa_1,...\pa_n;\ga_1,...\ga_m}\by_{\pa_1}\cdots \by_{\pa_n} y_{\ga_1}\cdots  y_{\ga_m}.
\ee
(See \cite{Vasiliev:1999ba} and references therein.)
In the $AdS_4$ background the parameter $\lambda$, that enters both the background
equations (\ref{nR}), (\ref{tors}) and free field equations (\ref{CON1})-(\ref{tw}),
is real and  determines a value and sign of the cosmological constant,
that correspond to $AdS_4$.

The extension to the star product (\ref{star2}) results from
\be
\label{dz}
(f(z,y,\bar z,\bar y))^\dagger  = \bar f (-\bar z,\bar y, - z, y)\,.
\ee
 So defined $\dagger$ forms an involution of the star product (\ref{star2}),
 $
(f* g)^\dagger = g^\dagger  * f^\dagger \,.
$
Its extension to Klein operators results from
\be
\label{dk}
k^\dagger = \bar k\q (\bar k)^\dagger = k\q
(k f(Z,Y))^\dagger = f^\dagger (Z,Y) \bar k \q (\bar k f(Z,Y))^\dagger = f^\dagger (Z,Y)  k\,.
\ee

An important property of  $\dagger$ (\ref{dz}), (\ref{dk}) is that
\be\label{ggadag}
\gga^\dagger = -\bar \gga\q \bar \gga^\dagger =- \gga \, \ee
taking into account that $(\theta^\ga \theta_\ga)^\dagger = - \bar \theta^\pa \bar \theta_\pa $
for $(\theta_\ga)^\dag =\bar \theta_\pa$.
It allows us to impose the following Hermiticity conditions on the
 fields of the $AdS_4$ HS system  (\ref{WW})-(\ref{SB})
 \be
\label{adsherm}
W^\dagger (Z,Y,K) = - W  (Z,Y,K)\,,\quad S^\dagger (Z,Y,K) =-  { S}(Z,Y,K)\,,\quad
B^\dagger (Z,Y,K) =   {B}(Z,Y,K)\,.
\ee

The reality conditions (\ref{adsherm}) correspond to the unitary HS theory in $AdS_4$. One can consider  more general reality conditions with involutions ${\dag'}$ (\ref{dag'}). If ${\dag'}$ leaves invariant
the original background fields bilinear in $Y$, then it still induces alternative reality conditions on the $AdS_4$ HS theory. The resulting $AdS_4$ HS theory is either equivalent or non-unitary.

\section{HS theory in $dS_4$}
\label{dS4 Hermiticity}
Since  the sign of the cosmological constant in $dS_4$ is opposite,
using the formulation invented for the $AdS_4$ HS theory,  one is forced to
demand $\lambda^2$ be negative. This is possible in the purely bosonic HS theory
with $\lambda$ entering via $\lambda^2$ but not in
presence of fermions in which case $\lambda$ contributes itself to the
mass-like terms in the field equations. This is clear from a simple dimension
argument: $\lambda$ is the only dimensionful parameter in the theory, $[\gl] = cm^{-1}$, and hence
the power of its  contribution to the mass-like terms in the field equations
is the same as the order of space-time derivatives, \ie
$\lambda^2$ for bosons and $\lambda$ for fermions (for explicit formulae see the
original papers \cite{Fronsdal:1978vb,Fang:1979hq}).
 As a result, supersymmetric HS
theory in $AdS_4$ becomes ill-defined in $dS_4$.  The main point of this paper is that it is possible to modify
the Hermiticity conditions in a way compatible with fermions without awkward factors of $i$.

\subsection{Type-I  theory}
\label{I}

To describe fields of all integer and half-integer spins
 in $dS_4$
 we start with  the following Hermiticity conjugation motivated by \cite{Vasiliev:1985vtv}, which we call type IA,
 \bee \label{HermdS}
 (i)^\ddag=-i\q k^\dgr=\bk\q \bk^\dgr= k\q (AB)^\dgr=B^\dgr A^\dgr\q (\dr_x)^\dgr=-\dr_x
\q\\ \nn
 {y}^\dgr_\alpha:=i   \by_\pa \KK  \q  {\by}^\dgr_\pa:=-i   {y}_\ga \KK \q
   {z}^\dgr_\alpha:=-i   \bz_\pa \KK \q  {\bz}^\dgr_\pa:= i   {z}_\ga \KK\,.
\eee
Since
$
(y_\ga y_\ga)^\ddag = \bar y_\pa \bar y_\pa\,, (\bar y_\pa\bar  y_\pa)^\ddag = y_\ga y_\ga\,
$
 the Lorentz connection components $\go_{\ga\ga}$ and $\bar \go_{\pa\pa }$ are
mutually conjugated as in the  $AdS_4$ case. However,
because of the Klein operator $\KK$ in (\ref{HermdS}),
$
(y_\ga \bar y_\pb )^\ddag = - y_\gb \bar y_\pa
$
and, hence,
\be\label{hdS}
\bar h_{\ga\pb} = - h_{\gb\pa}\,.
\ee

Since $\lambda$ in HS equations appears in combination with the vierbein $\lambda h_{\ga\pa}$ one might guess
that the $dS_4$ involution $\ddag$ is equivalent to the $AdS_4$ one $\dag$ trading imaginary $\lambda$ for imaginary
$h_{\ga\pa}$. This is indeed true for the bosonic HS theory but not for  fermions since $\ddag$
(\ref{HermdS}) mutiplies odd functions of $y_\ga$ and $\bar y_\pa$, \ie fermions, by the Klein operator $\KK$ thus mapping
 fermionic fields independent of $k$ and $\bar k$ to those proportional to $k\bar k$ and vice versa.\footnote{Recall that Fang-Fronsdal fields of spins $s>1/2$ belong to the one-form generating function $\go(Y,k,\bar k)$ even in $K$. Though spin 1/2 fields belong to the zero-form generating function $C(Y,k,\bar k)$ odd in $K$ the mass-like term is zero in that case as one can see from
 the spin-1/2 equation in (\ref{CON2}).}. In other words the conjugation (\ref{HermdS}) exchanges
the two types of fermionic fields in the  HS theory. It is this doubling, that allows one to avoid
the imaginary unit in the fermionic mass-like terms.

Since \be
 {\kappa}^\dgr=\bar{\kappa}\q\bar{\kappa}^\dgr={\kappa}\q (\theta^{\ga}\theta_{\ga})^\dgr =
        - \bar{\theta}^{\pa}\bar{\theta}_{\pa}\,\q
         (\bar{\theta}^{\pa}\bar{\theta}_{\pa})^\dgr =
         -\theta^{\ga}\theta_{\ga}\,,
  \ee
where $ {\kappa}$ and $\bar{\kappa}$ are left and right inner Klein operators
\eq{kk4},
 the central elements $\gga$ and $\bga$ \eq{gamma=} satisfy
\be\label{bilindaggamma}\gamma^\dgr =-\bga\q  \bga^\dgr=-\gga . \ee
Hence the involution $\ddag$ allows one to impose the reality conditions
analogous to
(\ref{adsherm}).
That these conditions are consistent with the full nonlinear system,
implies analogously to the $AdS_4$ case that the 
hermiticity conditions
\eq{HermdS} 
respect HS equations in all orders.

To present explicit formulae for the involution $\ddag$ in terms of the HS generating fuctions we observe that
from \eq{HermdS} it follows that for
 \be \label{DefF}  F^{a,b} (z,\bz,y,\by) *  k^a*\bk^b=  f {\,}^{\gl(m),\pl(l),\mu(n),\bmu (j), a,b }
   z_{\gl(m)} \bz_{\pl(l)}y_{\mu(n)} \by_{\bmu (j)} * k^{a }* \bk^{b}\,
\ee
\be \nn\big(F ^{a,b} (z,\bz,y,\by) *  k^a*\bk^b\big)^\dgr =
\bk^{a   }* k^{ b } *\overline{  f} {\,}^{\gl(m),\pl(l),\mu(n),\bmu (j), a,b }
\, (-i\bz\KK)_{\pl(m)}  (iz\KK)_{\gl(l)}(i\by\KK)_{\bmu (n)}  (-iy\KK)_{\mu(j)}
\ee
 \be\label{DefFdag}  = \bk^{a   }* k^{ b } *\overline{F}{} ^{a,b}
    (-\bz,  z , \by, - y )  (i \KK)^{\pi_F}\,.
\ee
where $\pi_F=0$ or 1 is a boson-fermion parity of ${F}$. We leave it to the reader as a simple exercise
to check that this formula describes an involution of the star product.

The construction of this section is somewhat analogous to that of \cite{LukNow2,PilchAtall} where the fermion
doubling was within extended SUGRA. Also it is analogous to the HS construction of Sezgin and Sundell
\cite{Sezgin:2012ag} with the important difference that our conjugation involves the outer Klein element $\KK$,
that anticommutes with  the differentials $\theta_A$, while the
conjugation of  \cite{Sezgin:2012ag} was defined in terms of inner Klein elements
built from the Clifford elements underlying the supersymmetric extension of the global space-time algebra \cite{Konstein:1989ij}.
That is fine at the free field level considered in  \cite{Konstein:1989ij} but may be problematic within the full nonlinear HS system.

Now we are in a position to define an alternative conjugation \eq{dag'}-\eq{rho},
referred to as type IB, that generalizes that of \cite{Lang:2026inn} to the case with fermions:
  \be\label{dddag}
 (C(\mathbf{Y}|x))^{\dddag}:=i^{\pi_C}    \rho\big[\big(C(\mathbf{Y}|x)\big)^{\ddag}\big].\ee

\subsection{Type-II theory  and  $\mathbb{Z}_2\times \mathbb{Z}_2$ structure}
\label{Z2Z2}
Let us introduce two sets of oscillators
 \be \label{oscdag}
\iy_\ga :=  {y}_\ga \q   \biy_\ga:= i   \by_\pa \KK \q
\iz_\ga:= {z}_\ga  \q   \biz_\ga:=-i   \bz_\pa \KK\,,
\ee
that obey the (anti)commutation relations
  \bee
\label{id4osca}
\ls&&[ \iy_\alpha ,  \iy_\beta ]_* = 2i  \epsilon_{\alpha\beta}, \quad
[ \biy_{\dot{\alpha}} ,  \biy_{\dot{\beta}} ]_* =  2i
\epsilon_{\dot{\alpha}\dot{\beta}}, \quad
[ \iz_\alpha ,  \iz_\beta ]_* = -2i  \epsilon_{\alpha\beta}, \quad
[ \biz_{\dot{\alpha}} ,  \biz_{\dot{\beta}} ]_* = - 2i
\epsilon_{\dot{\alpha}\dot{\beta}}, \qquad\\ \rule{0pt}{21pt}\nn
&& \quad
\{ \iy_\alpha ,  \biy_{\dot\beta} \}_* =0\q \{ \iz_\alpha ,  \biz_{\dot\beta} \}_* =0 \q \{ \iz_\alpha ,  \biy_{\dot\beta} \}_* =0\,\q \{ \iy_\alpha ,  \biz_{\dot\beta} \}_* =0\,.
\eee
Of course, the alternative version of (\ref{oscdag}) with  exchanged $y_\ga$, $z_\ga$ and $\bar y_\pa$, $\bar z_\pa$ is also possible.

Relations (\ref{id4osca}) exhibit a structure of $\mathbb{Z}_2\times \mathbb{Z}_2$
algebra \cite{rw1,rw2} endowed with the two $\mathbb{Z}_2$ gradings $\pi_1$ and $\pi_2$ and commutation relations
defined in terms of generalized superanticommutators
\be
\{\{A\,,B\}\} := A B + (-1)^{\pi_1(A) \pi_1(B)+\pi_2 (A) \pi_2 (B)} B A\,
\ee
and commutators
\be
[[A\,,B]] := A B - (-1)^{\pi_1(A) \pi_1(B)+\pi_2 (A) \pi_2 (B)} B A\,
\ee
 with
\be
\pi_1 (\iy_\ga,\iz_\ga)=1\q \pi_2 (\iy_\ga,\iz_\ga ) = 0\q \pi_1 (\biy_\pa,\biz_\pa)=0\q \pi_2 (\biy_\pa,\biz_\pa) = 1
\,.
\ee
 For instance, 
  $[[\iy_\ga\,,\iy_\gb]] =
  \{\iy_\ga\,,\iy_\gb\}\,,\, \{\{\iy_\ga\,,\,\iy_\gb\}\}
  =[\iy_\ga\,,\iy_\gb].$
For
$
\textrm{Y}_A:= (\iy_\ga,\biy_\pa)\,
$
relations 
(\ref{id4osca}) take the form
\be
\{\{\Y_A\,,\Y_B\}\} = 2i C_{AB}\q C_{AB} = (\epsilon_{\ga\gb},\epsilon_{\pa\pb} )\,.
\ee

Then (super)generators
\be\label{TAB=}
T_{AB}:=  [[\Y_A\,, \Y_B ]]\q
Q_A := \Y_A
\ee
form  the $\mathbb{Z}_2\times \mathbb{Z}_2$ generalized Lie superalgebra
counterpart of the usual superalgebra $osp(1,4)$,
introduced in \cite{Vasiliev:1985vtv} to
formulate $dS_4$ supergravity. Remarkably, both  $\mathbb{Z}_2$ and
$\mathbb{Z}_2\times \mathbb{Z}_2$ structures
result from the same associative HS algebra.

Note that now the $\mathbb{Z}_2\times \mathbb{Z}_2$ superalgebraic relations read as
\be
P_{\ga\pa} = [Q_\ga \,,\overline Q_\pb]\q
L_{\ga\gb} = \{\iy_\ga\,,\iy_\gb\}\q \bar L_{\pa \pb}  = \{\biy_\pa \,,\biy_\pb\}
\ee
with the commutator for $P_{\ga\pa}$ rather than anticommutator in
$osp(1,4)$.

The conjugation relations are
\be\label{conjII}
(\iy_\ga)^\ddag  = \biy_\pa\q
(\biy_\pa)^\ddag  = \iy_\ga\,,
\ee
\be
(L_{\ga\gb})^\ddag  = \bar L_{\pa\pb}\q (\bar L_{\pa\pb})^\ddag = L_{\ga\gb} \q (P_{\ga \pb})^\ddag = - P_{\gb \pa}\,.
\ee
The last relation again implies that though  $\lambda $ is real, the bosonic sector of the
$\mathbb{Z}_2\times \mathbb{Z}_2$ superalgebra in question
describes $dS_4$ rather than $AdS_4$ symmetry. The conjugation relations (\ref{conjII}) will be referred to as type IIA. As in type I theory,  type IIB reality conditions are defined by formula (\ref{dddag}) with
$\ddag$ (\ref{conjII}). Also note that the type-I and type-II $dS_4$ theories can be related by the replacement of the vierbein
$h_{\ga\pa}y^\ga\bar y^\pa$ by  $h_{\ga\pa}\KK y^\ga\bar y^\pa$.

In the standard $AdS_4$ HS theory massless fields in  the  zero-form sector
 belong to the twisted adjoint representation in which $P_{\ga\pb}$ acts by anticommutator because  the zero-forms  associated with the massless fields are odd in $k$
or $\bar k$, \ie $C(Y,-k,-\bar k) = - C(Y,k,\bar k)$.
 The fields of the opposite $k,\bar k$
parity belong to the adjoint representation  and are topological, each describing a
finite number of DoF \cite{Vasiliev:1987hv,Vasiliev:1988sa,Kirakosiants:2026spd}.

In the $\mathbb{Z}_2\times \mathbb{Z}_2$ realization of the HS algebra the situation is different.
Here the terms with $P_{\ga\pa}$ enter the covariant derivative through commutator with fermions odd in $k,\bar k$ and anticommutator
with fermions  even in $k,\bar k$.
This implies that the roles of the fermions of definite $k$, $\bar k$ parity
  are exchanged: $k$, $\bar k$ even fermions in the sector of zero-forms are massless while $k$, $\bar k$ odd fermions are topological. This suggests
the following speculation.

First of all let us stress that though any irreducible topological (massless) field
carries a finite (infinite) number of DoF, the full sets of massless and topological fields carry the same (infinite) numbers of DoF associated with unrestricted functions $C(Y)$.
The topological fields were interpreted in \cite{Didenko:2015pjo} as a set of coupling constants (moduli) of the HS theory.
The idea is that though it is impossible to interpolate from $AdS_4$ background to the $dS_4$ one in HS theory
staying within maximally symmetric backgrounds with unbroken HS symmetries since the latter are incompatible
with Minkowski background, there is a possibility to interpolate from $AdS_4$ to $dS_4$ via less symmetric
states with nontrivial topological fields, that necessarily break down maximal symmetries of the vacuum state.

 \section{Unitarity}
 \label{uni}

In this section we check which of the proposed systems respect unitarity and which do not. Since we are working with the field equations rather than the action we need an appropriate criterion at this level. To this end we will follow the approach of \cite{Gelfond:2013xt} to the computation of OPE current algebra in $3d$ and $4d$ HS theories. Namely, we consider Pauli-Jordan $D$-functions associated with the HS equations in question and check relative signs between their parts
 associated with different spins. To simplify the analysis we consider the flat limit of the HS equations since non-unitarity in the flat limit would imply that in $(A)dS$.

\subsection{Flat limit of the linearized $AdS_4$ HS theory}
The flat limit of the free equations in $AdS_4$ in the Cartesian coordinates with
zero Lorentz connection, $h^{\ga\pb}=  d x^{\ga\pb}$ and conventions \eq{invads} reads as
\be\label{fltwEqAdS}
 \big[ \ff{\p}{\p x^{\ga\pb}}   +i  \p_\ga \bbp_\pb\big]C(Y;K|x)=0\,.
\ee
 Equation  \eq{fltwEqAdS} admits  positive- and negative-frequency solutions, that can be  represented as
 \be \label{CflatAdSp} C_{{}_{AdS}}^+  (Y;K|x)=
   \int d^4\xi  \exp i\, [
  \xi_{\ga}\bar{\xi}_{\pb} x^{\ga\pb}  +        \gx_\ga y^\ga+       \bar{\xi}_\pa \by^\pa ]
 * \big[   k  F^+ _{1,0}( \gx  ,  {\bgx} )
 + \bk   F^+_{0,1}( \gx  ,  {\bgx} )  \big]\,,
 \ee
\be
 \label{CflatAdSm}
 C_{{}_{AdS}}^-  (Y;K|x)=
   \int d^4\xi \big[   k  F^- _{1,0}( \gx  ,  {\bgx} )
 + \bk  F^-_{0,1}( \gx  ,  {\bgx} )  \big] *\exp -i \,[ \,\xi_{\ga}\bar{\xi}_{\pb} x^{\ga\pb} +\gx_\ga y^\ga+\bar{\xi}_\pa \by^\pa ]
 \, .    \ee

The integration variables $\xi_\ga$ and $\bar\xi_\pa$ commute with
the Klein operators and are complex conjugated. Unrestricted functions
$F^\pm _{1,0}( \gx  ,  {\bgx} )$ and $ F^\pm_{0,1}( \gx  ,  {\bgx} )$ parameterize
the space of solutions. Since  $\xi_{\ga}\bar{\xi}_{\pb}$ is semi-positive definite,
the positive- (negative-) frequency solutions $C_{{}_{AdS}}^+$ $(C_{{}_{AdS}}^-)$
   have to be associated with the creation (annihilation) operators upon quantization.

Let us stress that the involution $\dagger$  maps equation (\ref{fltwEqAdS}) to itself since, though
$\overline{
\frac{\p}{\p x^{\ga\pa}}}=\frac{\p}{\p x^{\ga\pa}}$,
$\Big(
\frac{\p}{\p x^{\ga\pa}}\Big )^\dagger=-\frac{\p}{\p x^{\ga\pa}}\,.
$
The involution
$\dagger$ is designed to preserve the form of the nonlinear HS system and, hence, of its linearization.
(Recall that $(k *f(Y))^\dag = (f(Y))^\dag * \bar k $).

One can see that $(C_{{}_{AdS}}^\pm (Y;K|x) )^\dag=  \widetilde{C}_{{}_{AdS}}^\mp (Y;K|x)   $, where
\bee \label{CflatAdSddag+} &&   \widetilde{C}_{{}_{AdS}}^- (Y;K|x)=
  \int d^4\xi
      \big[  \bk  \bar{F}^+ _{1,0}( \bgx  ,  {\gx} )
 + k  \bar{F}^+_{0,1}( \bgx  ,  {\gx} )  \big] *  \exp
   - i\, [ \xi_{\gb}\bar{\xi}_{\pa} x^{\pa\gb} +
          \bar{\xi}_\pa \by^\pa +        \gx_\ga y^\ga]
    \,, \\  \label{CflatAdSddag-} &&\widetilde{C}_{{}_{AdS}}^+ (Y;K|x)= 
  \int d^4\xi  \exp
  i\, [ \xi_{\gb}\bar{\xi}_{\pa} x^{\pa\gb} +
          \bar{\xi}_\pa \by^\pa +        \gx_\ga y^\ga]
     * \big[  \bk  \bar{F}^- _{1,0}( \bgx  ,  {\gx} )
 + k  \bar{F}^-_{0,1}( \bgx  ,  {\gx} )  \big]
   \,. \eee

  Reality conditions yield $ C^\pm =\widetilde{C}^\pm$\,. Hence
 \bee \label{AdSReal}&&     F^+_{0,1}( \gx  ,  {\bgx})= \bar{F}^- _{1,0}( \bgx  ,  {\gx}   ) \q
 F^+ _{1,0}( \gx  ,  {\bgx})= \bar{F}^-_{0,1}( \bgx  ,  {\gx} )
   \,. \eee

It is sometimes convenient to perform integration with respect to $\gx,\bgx$.
To this end,  following \cite{Gelfond:2008ur}, we introduce
\be\label{Sieg}   \Z_{\pm}^{\ga\pb}
=  x^{\ga\pb}\pm  i e^{\ga\pb} \ee
with an infinitesimal real positive-definite matrix $e^{\ga\pb}$ making the
Gaussian integrals  \eq{CflatAdSp}, \eq{CflatAdSm} as well as
    $\det(  \pm i \Z^{\ga\pb}_\pm){}^{-\half}  $  well defined (for more details see \cite{Gelfond:2008ur}).
   Denoting  $  \p_\ga:=\ff{\p}{\p  y^\ga}\, $,
$  \bar\p_\pa :=\ff{\p}{\p  \by^\pa}\, $
and evaluating the Gaussian integrals one obtains from \eq{CflatAdSp}, \eq{CflatAdSm}
  \bee \label{CflatAdSpGa}\ls &&C^+  (Y;K|x)=
  \pi
 \big(\det (  -i \Z_{+} )\big)^{ -\half}    \exp (  -i(\Z_+)^{-1}_{\ga \pb } y^\ga \by^\pb )
    \big[   k F^+_{1,0}(  -     \overleftarrow{i\p} , -   \overleftarrow{i\bar\p} )+
\bk  F^+_{0,1}(  -    \overleftarrow{i\p} , - \overleftarrow{i\bar\p} )
  \big] \q\\ \label{CflatAdSmGa}\ls &&C^-  (Y;K|x)=
  \pi   \big[     k F^-_{1,0}(    i  {\p} ,   i  {\bar\p} )+
\bk  F^-_{0,1}(    i  {\p} ,  i   {\bar\p} ) \big]\big(\det (  i  {\Z_{-}})\big)^{ -\half}
   \exp (  i(\Z_-)^{-1}_{\ga \pb } y^\ga \by^\pb )
 \, .    \eee

Hence   \bee \label{CflatAdSpGad}\ls\! &&(C^+  (Y;K|x))^\dag\!=\!
  \pi\!
   \big[   \bk \overline{F}^+_{1,0}( {i\bar\p}, {i\p}    )+
 k  \overline{F}^+_{0,1}(  {i\bar\p}, {i\p}     )
  \big(\det (  i\,\overline{ \Z_{+}} )\big)^{ -\half}    \exp ( i(\overline{\Z_+})^{-1}_{\pa \gb } y^\gb \by^\pa )
   \big] \q\,\,\\ \label{CflatAdSmGad}\ls\! &&(C^-  (Y;K|x))^\dag
   =\pi \big(\det (  - i \, {\overline{\Z_{-}}})\big)^{ -\half}
   \exp (  -i(\overline{\Z_-})^{-1}_{\ga \pb } y^\ga \by^\pb )
 \big[   \bk \overline{F}^-_{1,0}(- \overleftarrow{i\bar\p} ,-\overleftarrow{i\p})+
 k \overline{ F}^-_{0,1}(- \overleftarrow{i\bar\p} ,-\overleftarrow{i\p}  )\big] .\quad\qquad    \eee

The Pauli-Jordan $D$-functions, that determine the form of the two-point correlator can be represented as the product of positive- and negative- frequency
solutions
\be
D(Y,Y'|x,x') = C^-(Y|x) C^+(Y'|x')
\ee
with particular coefficient functions $F^\pm$ in (\ref{CflatAdSp}), (\ref{CflatAdSm}). (For more detail see \cite{Gelfond:2008ur,Gelfond:2013xt}.)
Hence the relative signs of the positive- and negative-frequency solutions for different spins determine relative signs of the two-point correlators that distinguish between unitary and nonunitary models. The unitary $AdS_4$ HS model  serves as the reference model that determines proper relative signs compatible with unitarity.

 \subsection{Flat limit of the linearized type-I $dS_4$   HS theory   }
\subsubsection{Type-IA}
\label{IA}
Using \eq{HermdS}, \eq{hdS} and setting
\be\label{hidx}
 h^{\ga\pb}=i d x^{\ga\pb}
\ee
one obtains from \eq{CON2}, \eq{fltw}
\be\label{fltwEq=dS}
 \big[ \ff{\p}{\p x^{\ga\pb}}   -  \p_\ga \bbp_\pb\big]C(Y;K|x)=0\,.
\ee

With \be\label{bfyI}\mathbf{y}_\ga:=-i y_\ga\q \mathbf{\by}_\pa:=\by_\pa
\ee
equation \eq{fltwEq=dS} acquires the form
 \be\label{fltwEqbfI}
 \big[ \ff{\p}{\p x^{\ga\pb}}  +  i\ff{\p^2}{\p \mathbf{y}^\ga \p \mathbf{\by}^\ga} \big]C'(\mathbf{y}, \mathbf{\by};K|x)=0\q C'(\mathbf{y}, \mathbf{\by};K|x):= C(i\mathbf{y}, \mathbf{\by};K|x) \,.
\ee  Hence  $ C^\pm  (Y|x)$   \eq{CflatAdSp}, \eq{CflatAdSm}  with $y,\by\to \mathbf{y}, \mathbf{\by} $ solves \eq{fltwEqbfI}:
 \bee \label{CflatdSp}\ls\ls &&C'{}^+  (\mathbf{Y};K|x) =
   \int d^4\xi  \exp i\, [
  \xi_{\ga}\bar{\xi}_{\pb} x^{\ga\pb}  +        \gx_\ga \mathbf{y}^\ga+       \bar{\xi}_\pa \mathbf{\by}^\pa ]
  \big[   k  F^+ _{1,0}( \gx  ,  {\bgx} )
 + \bk   F^+_{0,1}( \gx  ,  {\bgx} )  \big],\,
 \\ \label{CflatdSm} \ls\ls&& C'{}^-(\mathbf{Y};K|x)  =
   \int d^4\xi \big[   k  F^- _{1,0}( \gx  ,  {\bgx} )
 + \bk  F^-_{0,1}( \gx  ,  {\bgx} )  \big] \exp -i \,[ \,\xi_{\ga}\bar{\xi}_{\pb} x^{\ga\pb} +\gx_\ga \mathbf{y}^\ga+\bar{\xi}_\pa \mathbf{\by}^\pa ]
  .  \,  \eee
Then, using that, by virtue
\eq{HermdS}, \eq{bfyI},
   \be\label{dddagy}
    \mathbf{y}^\ddag= 
    - \mathbf{\by}\KK\,,
 \, \mathbf{\by}^\ddag  
 =\mathbf{y} \KK\,,
 \ee
 one finds with the aid of \eq{DefFdag}
 \bee \label{CflatdSpdag=} &&(C'{}^+  (\mathbf{Y};K|x))^\ddag =
   \int d^4\xi\exp \, [-i
  \xi_{\ga}\bar{\xi}_{\pb} x^{\ga\pb}] \big[    \bk  \overline{F}^+ _{1,0}( \bgx  ,  {\gx} )
 +  k   \overline{F}^+_{0,1}( \bgx  ,  {\gx} )  \big] \times   \\ \nn&&\times\big\{
 \sinh[ -       \gx_\ga \mathbf{y}^\ga + \bar{\xi}_\pa \mathbf{\by}^\pa  ] i\KK
  +    \cosh[ -       \gx_\ga \mathbf{y}^\ga + \bar{\xi}_\pa \mathbf{\by}^\pa  ]
  \big\}\,,
 \eee
 \bee \label{CflatdSmdag=} &&(C'{}^-(\mathbf{Y};K|x))^\ddag =
   \int d^4\xi  \exp    \,[ i\,\xi_{\ga}\bar{\xi}_{\pb} x^{\ga\pb}] \big\{
   \sinh[          \gx_\ga \mathbf{y}^\ga -\bar{\xi}_\pa \mathbf{\by}^\pa  ]i\KK
  +\\ \nn&& +\cosh[        \gx_\ga \mathbf{y}^\ga-\bar{\xi}_\pa \mathbf{\by}^\pa ]
  \big\}
  \big[   \bk  \overline{F}^- _{1,0}( \bgx  ,  {\gx} )
 +  k   \overline{F}^-_{0,1}( \bgx  ,  {\gx} )  \big]\, .    \eee

Comparing these formulae with the $AdS_4$ ones (\ref{CflatAdSddag+}) and
(\ref{CflatAdSddag-}) we observe that the trigonometric $\cos$ and $\sin$ are replaced by the hyperbolic $\cosh$ and $\sinh$ that differ in particular by the sign of the coefficients at odd $n$ in  the expansions
\be
\label{co}
\cos(x)= \sum_{n=0}^\infty (-1)^n \frac{1}{(2n)!} x^{2n} \q
\cosh(x) = \sum_{n=0}^\infty  \frac{1}{(2n)!} x^{2n} \,.
\ee
This implies that the relative signs between the $D$-functions (two-point correlators) of even and odd spins are different in the $AdS_4$ and type-IA $dS_4$ cases. Hence, in agreement  with the conclusions of \cite{PilchAtall}, the Hermitian conjugation of this
section leads to ghosts in the $dS_4$ HS theory, usually attributed to odd spins.  Note that the relative signs  of spins $s=1/2+4n$ and $s=3/2 +4n$  also differ from those in the unitary $AdS_4$ case.

  \subsubsection{Type-IB}
\label{alternativeI}

 From \eq{dddag} it follows that
\bee \label{CflatdSpdddag=} &&(C'{}^+  (\mathbf{Y};K|x))^\dddag=
   \int d^4\xi   \exp -i\, [
  \xi_{\ga}\bar{\xi}_{\pb} x^{\ga\pb}]
   \big\{  \big[   \bk   \overline{F}^+_{0,1}( \bgx  ,  {\gx} )
 -   k  \overline{F}^+_{1,0}( \bgx  ,  {\gx} ) \big]
    \sin[       \gx_\ga \mathbf{y}^\ga +  \bar{\xi}_\pa \mathbf{\by}^\pa  ]
  + \nn\\ && \qquad\qquad\qquad\qquad+\big[   \bk  \overline{F}^+ _{1,0}( \bgx  ,  {\gx} )
 +  k   \overline{F}^+_{0,1}( \bgx  ,  {\gx} )  \big]\cos[       \gx_\ga \mathbf{y}^\ga+\bar{\xi}_\pa \mathbf{\by}^\pa  ]
  \big\} \,,
 \eee
 \bee \label{CflatdSmdddag=} &&(C'{}^-(\mathbf{Y};K|x))^\dddag=
   \int d^4\xi  \exp  i \,[ \,\xi_{\ga}\bar{\xi}_{\pb} x^{\ga\pb}]
     \big\{
    \sin[          \gx_\ga \mathbf{y}^\ga+\bar{\xi}_\pa \mathbf{\by}^\pa  ]   \big[ \bk   \overline{F}^-_{0,1}( \bgx  ,  {\gx} )   -k  \overline{F}^- _{1,0}( \bgx  ,  {\gx} )
   \big]
  +\nn \\  &&\qquad\qquad\qquad\qquad + \cos[        \gx_\ga \mathbf{y}^\ga +\bar{\xi}_\pa \mathbf{\by}^\pa ]\big[   \bk  \overline{F}^- _{1,0}( \bgx  ,  {\gx} )
 +  k   \overline{F}^-_{0,1}( \bgx  ,  {\gx} )  \big]
  \big\}   .    \eee

Now the relative signs
 of the conjugated $D$-functions of integer spins associated with the $\cos$ terms coincide with those of the $AdS_4$ case.
This agrees with the recent results of Lang and Neiman \cite{Lang:2026inn}
who in fact used the same conjugation in the bosonic $dS_4$ HS theory.

The situation with fermions is different. It turns out that $\dddag$ is a pseudo involution in the sense of (\ref{pinv}). Indeed,
using \eq{DefFdag}, \eq{dddagy} and \eq{dddag}, it is not hard to see that
\be \label{CflatdSpdddag2} \big((C'{}^+  (\mathbf{Y};K|x))^\dddag\big)^\dddag
  =   \int d^4\xi \exp  i\, [
  \xi_{\ga}\bar{\xi}_{\pb} x^{\ga\pb}
       -   \gx_\ga \mathbf{y}^\ga -     \bar{\xi}_\pa \mathbf{\by}^\pa   ]
    \big[    k   {F}^+ _{1,0}( \gx  ,  {\bgx} )
 +  \bk   {F}^+_{0,1}( \gx  ,  {\bgx} )  \big]\,,
 \ee
 \be \label{CflatdSmdddag2} \big(( C'{}^-(\mathbf{Y};K|x))^\dddag\big)^\dddag  =
   \int d^4\xi \big[   k  F^- _{1,0}( \gx  ,  {\bgx} )
 + \bk  F^-_{0,1}( \gx  ,  {\bgx} )  \big] *\exp -i \,[ \,\xi_{\ga}\bar{\xi}_{\pb} x^{\ga\pb}-
 \gx_\ga \mathbf{y}^\ga-\bar{\xi}_\pa \mathbf{\by}^\pa ]
  .  \,  \ee

 As a result, the double application of $\dddag$ changes a sign of odd functions in $\mathbf{Y}$,
 \ie fermions, not affecting a sign of even functions, \ie bosons.

 \subsection{Flat limit of the linearized type-II $dS_4$  HS theory   }
\label{FLimII}

\subsubsection{Type-IIA}
 \label{FLimIIA}

 From \eq{oscdag}, \eq{conjII} it follows   \be \label{conjII=}
( y)^\ddag  =i   \by_\pa \KK \q
(  \by_\pa )^\ddag  = -i y_\ga \KK \, .
\ee
  Using $(y_\ga,\by_\pa)$ variables instead of $( \iy_\ga \,,\biy_\pa)$
and setting
$h^{\ga\pb}=i d x^{\ga\pb}
$
as in (\ref{hidx}),
one obtains from  \eq{fltw}
\be\label{CflatdS II}
 \big[ \ff{\p}{\p x^{\ga\pb}}   - \KK  \ff{\p^2}{\p y^{\ga }\p \by^{ \pb}}\big]C (Y;K|x)=0\,
\ee
and hence
 \be\label{fltwEqbfII}
 \big[ \ff{\p}{\p x^{\ga\pb}}  +  i\ff{\p^2}{\p \mathbf{y}^\ga \p \mathbf{\by}^\ga}\KK \big]
 C'(\mathbf{y}, \mathbf{\by};K|x)=0\q C'(\mathbf{y}, \mathbf{\by}):= C(i\mathbf{y}, \mathbf{\by};K|x) \,.
\ee

 Taking into account that, as   mentioned   in Section \ref{Z2Z2},  any even in $\mathbf{Y}$ solution  to \eq{CflatdS II} $C'$ (boson) is odd in $k,\bar k$
while any odd in $\mathbf{Y}$ solution  (fermion) is even in $k,\bar k$,
one can see that a general odd (even) solution $C'_o$ ($C'_e$)
to  \eq{fltwEqbfII} has  the form
 \bee \label{OddCII}
C'_o (\mathbf{Y} ;K|x)= i \int d^4\xi  \exp i\, [
  \xi_{\ga}\bar{\xi}_{\pb} x^{\ga\pb}]\ls&&\ls\big\{
  \Pi_+^0   \sin[\gx_\ga \mathbf{y}^\ga+   \bar{\xi}_\pa \mathbf{\by}^\pa ]
  F^1_+( \gx  ,  {\bgx} )
 +\\\nn&+& \Pi_-^0
\sin[ \gx_\ga \mathbf{y}^\ga-   \bar{\xi}_\pa \mathbf{\by}^\pa ]  F^1_-( \gx  ,  {\bgx} )
    \big\}     +\\ \nn+
 i \int d^4\xi  \exp -i\, [
  \xi_{\ga}\bar{\xi}_{\pb} x^{\ga\pb}]\ls&&\ls\big\{
   \Pi_-^0  G^1_-( \gx  ,  {\bgx} )\sin[ \gx_\ga \mathbf{y}^\ga+  \bar{\xi}_\pa \mathbf{\by}^\pa ]
 +\\\nn&+& \Pi_+^0  G^1_+( \gx  ,  {\bgx} )
\sin[\gx_\ga \mathbf{y}^\ga-\bar{\xi}_\pa \mathbf{\by}^\pa ]     \big\}   \,,   \eee
 \bee \label{EvenCII}
C'_e (\mathbf{Y} ;K|x)=  \int d^4\xi  \exp i\, [
  \xi_{\ga}\bar{\xi}_{\pb} x^{\ga\pb}]\ls&&\ls\big\{
  \cos[\gx_\ga \mathbf{y}^\ga+   \bar{\xi}_\pa \mathbf{\by}^\pa ]
   \Pi_+^1  F^0_+( \gx  ,  {\bgx} )
 +\\\nn&+&
 \Pi_-^1 \cos[  \gx_\ga \mathbf{y}^\ga+   \bar{\xi}_\pa \mathbf{\by}^\pa ]
    F^0_-( \gx  ,  {\bgx} ) \big\}     +\\ \nn+
  \int d^4\xi  \exp -i\, [
  \xi_{\ga}\bar{\xi}_{\pb} x^{\ga\pb}]\ls&&\ls\big\{
    \cos[ \gx_\ga \mathbf{y}^\ga+   \bar{\xi}_\pa \mathbf{\by}^\pa ] \Pi_-^1  G^0_-( \gx  ,  {\bgx} )
   +\\\nn&+&\Pi_+^1
\cos[ \gx_\ga \mathbf{y}^\ga+   \bar{\xi}_\pa \mathbf{\by}^\pa ]G^0_+( \gx  ,  {\bgx} )
    \big\}\,\q     \eee
where\be \Pi_\pm^0:=\half( 1 \pm\KK)\q\Pi_\pm^1:=\half( k \pm\bk)\q
\Pi_\pm^i \Pi_\pm^j = \Pi_\pm^{(i+j)|_{mod_2}}\q \Pi_-^i\Pi_+^j =0\,.
 \ee

Since the $\Pi_+^j$ proportional terms $C'{}^+_o$ of \eq{OddCII} and  $C'{}^+_e$ of \eq{EvenCII} solve
AdS-like equation  (\ref{fltwEqbfI})
while the other terms do not, we
consider these parts separately. Introducing \be \label{bfprim}\mathbf{y}'=\mathbf{y}\q \mathbf{\by}'=-\mathbf{\by},\ee
 one can see that  the following $C'{}^-_o (\mathbf{Y}' ;K|x)$  and $C'{}^-_e (\mathbf{Y}' ;K|x)$ solve
  \eq{fltwEqbfI} (with $\mathbf{Y} \to \mathbf{Y}'$)
  \bee \label{OddCII-}
C'{}^-_o (\mathbf{Y'}{} ;K|x)= i \int d^4\xi  \exp i\, [
  \xi_{\ga}\bar{\xi}_{\pb} x^{\ga\pb}]\ls&&\!
    \Pi_-^0 \sin[\gx_\ga \mathbf{y'}{}^\ga+   \bar{\xi}_\pa \mathbf{\by'}{}^\pa ]
    F^1_-( \gx  ,  {\bgx} )
     +\\ \nn+
 i \int d^4\xi  \exp -i\, [
  \xi_{\ga}\bar{\xi}_{\pb} x^{\ga\pb}]\ls&&\!
   \Pi_-^0  \sin[ \gx_\ga \mathbf{y'}{}^\ga-   \bar{\xi}_\pa \mathbf{\by'}{}^\pa ]G^1_-( \gx  ,  {\bgx} )
   \,,   \eee
  \bee \label{EvenCII-}
C'{}^-_e (\mathbf{Y'}{} ;K|x)=  \int d^4\xi  \exp i\, [
  \xi_{\ga}\bar{\xi}_{\pb} x^{\ga\pb}]\ls&&\!
   \Pi_-^1 \cos[  \gx_\ga \mathbf{y'}{}^\ga-   \bar{\xi}_\pa \mathbf{\by'}{}^\pa ]
     F^0_-( \gx  ,  {\bgx} )      +\\ \nn+
  \int d^4\xi  \exp -i\, [
  \xi_{\ga}\bar{\xi}_{\pb} x^{\ga\pb}]\ls&&\!
   \cos[ \gx_\ga \mathbf{y'}{}^\ga-  \bar{\xi}_\pa \mathbf{\by'}{}^\pa ] \Pi_-^1  G^0_-( \gx  ,  {\bgx} )
       \,.     \eee
  Now we are in a position to conjugate
$C'{}^-_o(\mathbf{Y'}{} ;K|x)$, $C'{}^-_e(\mathbf{Y'}{} ;K|x)$
and
$C'{}^+_o(\mathbf{Y}{} ;K|x)$, $C'{}^+_e(\mathbf{Y}{} ;K|x)\,.$

     Using \eq{bfyI}, \eq{conjII=},
 \eq{DefFdag}
 one finds
\bee \label{EvenCIIdd}
\big(C'{}^+_e (\mathbf{Y} ;K|x)\big)^\ddag=  \int d^4\xi  \exp -i\, [
  \xi_{\ga}\bar{\xi}_{\pb} x^{\ga\pb}]
  \ls&&\!
   \overline{\Pi_+^1  F^0}_+( \bgx  ,  {\gx} )\cosh[ \gx_\ga \mathbf{y}^\ga- \bar{\xi}_\pa \mathbf{\by}^\pa ]
      +\\ \nn+
  \int d^4\xi  \exp  i\, [
  \xi_{\ga}\bar{\xi}_{\pb} x^{\ga\pb}]
  \ls&&\!\cosh[  \gx_\ga \mathbf{y}^\ga- \bar{\xi}_\pa \mathbf{\by}^\pa ]
   \overline{\Pi_+^1 G^0}_+( \bgx  ,  {\gx} )
 \q     \eee
    \bee \label{OddCIIdd=}
\big(C'{}^+_o  (\mathbf{Y} ;K|x)\big)^\ddag= -i \int d^4\xi  \exp- i\, [
  \xi_{\ga}\bar{\xi}_{\pb} x^{\ga\pb}]\ls&&\!
   \Pi_ -^0 * \sinh[\gx_\ga \mathbf{y}^\ga- \bar{\xi}_\pa \mathbf{\by}^\pa]
    \overline{F^1}_+( \bgx  ,  {\gx} )        +
    \\ \nn+i
  \int d^4\xi  \exp  i\, [
  \xi_{\ga}\bar{\xi}_{\pb} x^{\ga\pb}]\ls&&\!  \Pi_ -^0
    \sinh[ \gx_\ga \mathbf{y}^\ga+\bar{\xi}_\pa \mathbf{\by}^\pa ]
      \overline{G^1}_+( \bgx  ,  {\gx} )\,.     \eee
 Using \eq{bfyI}, \eq{conjII=},
 \eq{DefFdag} and \eq{bfprim} one has
     $
    (\mathbf{y}')^\ddag= \mathbf{\by'}\KK     \,,
 \, (\mathbf{\by'})^\ddag
 =-\mathbf{y}' \KK\,,
  $  \bee \label{OddCII-dd}
\big(C'{}^-_o (\mathbf{Y'}{} ;K|x)\big)^\ddag=  -i \int d^4\xi  \exp- i\, [
  \xi_{\ga}\bar{\xi}_{\pb} x^{\ga\pb}]\ls&&\!  {\Pi_ +^0}
   \sinh[- \bgx_\pa \mathbf{\by'}{}^\pa   +  {\xi}_\ga \mathbf{ y'}{}^\ga ]
   \overline{ F^1_-}( \bgx  ,  {\gx} )
     +\\ \nn+ i
  \int d^4\xi  \exp  i\, [
  \xi_{\ga}\bar{\xi}_{\pb} x^{\ga\pb}]\ls&&\! {\Pi_ +^0}
\sinh[ \bgx_\pa \mathbf{\by'}{}^\pa   + {\xi}_\ga \mathbf{ y'}{}^\ga ]     \overline{ G^1_-}( \bgx  ,  {\gx} )\,,
\eee
   \bee \label{EvenCII-dd}
\big(C'{}^-_e (\mathbf{Y'}{} ;K|x)\big)^\ddag=   \int d^4\xi  \exp -i\, [
  \xi_{\ga}\bar{\xi}_{\pb} x^{\ga\pb}]\ls&&\!\big\{
   \overline{ \Pi_-^1  F^0}_-( \bgx  ,  {\gx}) \cosh[- \bgx_\pa \mathbf{\by'}{}^\pa   +  {\xi}_\ga \mathbf{ y'}{}^\ga ]
      \big\}     +\\ \nn+
   \int d^4\xi  \exp  i\, [
  \xi_{\ga}\bar{\xi}_{\pb} x^{\ga\pb}]\ls&&\!
    \cosh[- \bgx_\pa \mathbf{\by'}{}^\pa   +  {\xi}_\ga \mathbf{ y'}{}^\ga ]
       \overline{{ \Pi_-^1 }  G^0}_-(  \bgx  ,  {\gx} )\,.     \eee

One can see that, as in  Section \ref{IA}, the difference
between  $C'{}^-_{...}(\mathbf{Y'}{} ;K|x) $ ( $
C'{}^+_{...}(\mathbf{Y}{} ;K|x) \, $)
  and $(C'{}^-_{...}(\mathbf{Y'}{} ;K|x) )^\ddag$ ($(C'{}^+_{...}(\mathbf{Y }{} ;K|x) )^\ddag$),  is that trigonometric $\cos$ and $\sin$
functions are replaced by the hyperbolic ones $\cosh$ and $\sinh$, thus implying
non-unitarity.

  \subsubsection{Type-IIB}
 \label{alternativeII}
 From \eq{dddag} it follows that
\bee \label{EvenCIIddd}
\big(C'{}^+_e (\mathbf{Y} ;K|x)\big)^\dddag=  \int d^4\xi  \exp -i\, [
  \xi_{\ga}\bar{\xi}_{\pb} x^{\ga\pb}]
  \ls&&\!
    \overline{\Pi_+^1  F^0}_+( \bgx  ,  {\gx} )  \cos[ \gx_\ga \mathbf{y}^\ga+ \bar{\xi}_\pa \mathbf{\by}^\pa ]
      +\\ \nn+
  \int d^4\xi  \exp  i\, [
  \xi_{\ga}\bar{\xi}_{\pb} x^{\ga\pb}]
  \ls&&\!  \cos [  \gx_\ga \mathbf{y}^\ga+ \bar{\xi}_\pa \mathbf{\by}^\pa ]\overline{\Pi_+^1 G^0}_+( \bgx  ,  {\gx} )
 \q     \eee
   \bee \label{OdddCIIddd=}
\big(C'{}^+_o  (\mathbf{Y} ;K|x)\big)^\dddag=  \int d^4\xi  \exp- i\, [
  \xi_{\ga}\bar{\xi}_{\pb} x^{\ga\pb}]\ls&&\!
    \sin [-\gx_\ga \mathbf{y}^\ga+ \bar{\xi}_\pa \mathbf{\by}^\pa] \Pi_ -^0
    \overline{F^1}_+( \bgx  ,  {\gx} )   \big\}     +
    \\ \nn +
  \int d^4\xi  \exp  i\, [
  \xi_{\ga}\bar{\xi}_{\pb} x^{\ga\pb}]\ls&&\!     \sin [\gx_\ga \mathbf{y}^\ga+\bar{\xi}_\pa \mathbf{\by}^\pa ]
      \Pi_ -^0 \overline{G^1}_+( \bgx  ,  {\gx} ) \,.     \eee
   \bee \label{EvenCII-ddd}
\big(C'{}^-_e (\mathbf{Y'}{} ;K|x)\big)^\dddag=   \int d^4\xi  \exp -i\, [
  \xi_{\ga}\bar{\xi}_{\pb} x^{\ga\pb}]\ls&&\!\big\{
   \cos[{\xi}_\ga \mathbf{ y'}{}^\ga  - \bgx_\pa \mathbf{\by'}{}^\pa   ] \overline{{\Pi_-^1}
     F^0}_-( \bgx  ,  {\gx}) \big\}     +\\ \nn+
  \int d^4\xi  \exp  i\, [
  \xi_{\ga}\bar{\xi}_{\pb} x^{\ga\pb}]\ls&&\!\big\{
    \overline{ \Pi_-^1  G^0}_-(  \bgx  ,  {\gx} ) \cos[{\xi}_\ga \mathbf{ y'}{}^\ga - \bgx_\pa \mathbf{\by'}{}^\pa      ]
       \big\}\,.     \eee
   \bee \label{OdddCII-ddd=}
\big(C'{}^-_o (\mathbf{Y'}{} ;K|x)\big)^\dddag=  \int d^4\xi  \exp- i\, [
  \xi_{\ga}\bar{\xi}_{\pb} x^{\ga\pb}]\ls&&\!
    \sin [  \bgx_\pa \mathbf{\by'}{}^\pa   -  {\xi}_\ga \mathbf{ y'}{}^\ga ]{\Pi_ +^0}
   \overline{ F^1_-}( \bgx  ,  {\gx} )
     +\\ \nn +
  \int d^4\xi  \exp  i\, [
  \xi_{\ga}\bar{\xi}_{\pb} x^{\ga\pb}]\ls&&\!
  \sin [  \bgx_\pa \mathbf{\by'}{}^\pa   +  {\xi}_\ga \mathbf{ y'}{}^\ga ]
  {\Pi_+^0 }\overline{ G^1_-}( \bgx  ,  {\gx} )     \,.   \eee

Considering the terms proportional to $\Pi_+^1$  and $\Pi_-^1$ separately,
one finds that, as in Section \ref{alternativeI},
  the relative conjugation signs of the $D$-functions of integer spins associated with the $\cos$ terms are the same as in the $AdS_4$ theory thus suggesting
  unitarity of the bosonic sector.

 One can make sure that
 \bee \big(\big(C'{}^+_o (\mathbf{Y}{} ;K|x)\big)^\dddag\big)^\dddag=-C'{}^+_o (\mathbf{Y}{} ;K|x)\q
\big(\big(C'{}^-_o (\mathbf{Y'}{} ;K|x)\big)^\dddag\big)^\dddag=-C'{}^-_o (\mathbf{Y'}{} ;K|x)\q\eee
 \bee \big(\big(C'{}^+_e (\mathbf{Y}{} ;K|x)\big)^\dddag\big)^\dddag= C'{}^+_e (\mathbf{Y}{} ;K|x)\q
\big(\big(C'{}^-_e (\mathbf{Y'}{} ;K|x)\big)^\dddag\big)^\dddag= C'{}^-_e (\mathbf{Y'}{} ;K|x)\,.\quad\eee

 As a result, the double application of $\dddag$ changes a sign of odd functions in $\mathbf{Y}$,
 \ie fermions, not affecting that of even functions, \ie bosons.
 Thus $\dddag$ is a pseudo involution.

\section{Conclusion}
\label{conc}
Modified Hermiticity conditions  in the $dS_4$
HS gauge theory compatible with the presence of fermions are proposed that allow one to avoid problems with imaginary fermionic mass-like terms. Two of these types of Hermiticity conditions, that generalize
those proposed recently in \cite{Lang:2026inn} to the theory with fermions, lead to  pseudo involutions of the HS theories in $dS_4$. The same time, our
construction for  non-unitary nonlinear  HS theories with fermions is different from that proposed earlier in \cite{Sezgin:2012ag} without
the total Klein operator anticommuting with the differentials $dz^\ga=\theta^\ga$, which plays the key role in our approach.

It is speculated that the proposed mechanism of realization of the  $dS$ vacuum in HS theory may result from a tunneling process from $AdS$ vacuum to the $dS$ one in the nonlinear HS theory with topological fields since, at least in the one of the proposed models, the interpretation of the dynamical and topological fermionic fields is exchanged upon changing the $AdS$ reality conditions to the $dS$ ones.

A somewhat surprising observation of this paper is that the structure of $\mathbb{Z}_2\times \mathbb{Z}_2$ Lie superalgebra introduced in \cite{Vasiliev:1985vtv} for the description of $dS_4$ SUGRA admits a natural
realization in terms of the full nonlinear HS system of \cite{Vasiliev:1992av} which in turn was formulated with the help of Klein operators first found a field-theoretic application in the $dS_4$ SUGRA in
 \cite{Vasiliev:1985vtv}.

We hope that the construction of this paper may be useful in the
context of $dS_4$ HS/CFT holography and, more speculatively, the dark energy analysis.
A standard proposal for $dS_4$ HS/CFT holography is a nonunitary $sp(2N)$ or $usp(N,N)$  analogue  \cite{Anninos:2011ui}
 of the $AdS_4$ Klebanov-Polyakov model \cite{Klebanov:2002ja}
  with boundary anticommuting scalars $\phi^i (\xx)$
 and commuting spinors $\psi^i_\ga (\xx)$,
that commute with themselves,
$
\phi^i \psi_\ga^j = \psi_\ga^j \phi^i\,.
$
To agree with the $\mathbb{Z}_2\times \mathbb{Z}_2$ structure of the
type-II  $dS_4$ HS theory, it seems natural to speculated that
the commutation relations between $\phi^i$ and $\psi_\ga^j$  can be of the parastatistics type
  $
\phi^i \psi_\ga^j = - \psi_\ga^j \phi^i\,.
$

Direct analysis of $dS/CFT$ holography can be performed along the lines of the approach of \cite{Vasiliev:2012vf,Gelfond:2013xt}. We leave this interesting project for the future just mentioning that the pseudo involution of the proposed $dS_4$ HS models may be directly related to $dS_4$ holography since the boundary of $dS_4$ consists of two $S_3$ at opposite infinite times while the time reversal transform is antiunitary. A related fact is that fermions in the $3d$ Euclidean space posses pseudo involution rather than involution.

The doubling  of the boundary may be related to the construction of the involution from a pseudo involution in the doubled space presented in Section \ref{inv}.
A possibly related fact is that, as argued in \cite{Vasiliev:2012vf}, $AdS/CFT$
holography in presence of fermions suggests a doubling of the $AdS$ space upon crossing the boundary. Application of the analysis of \cite{Vasiliev:2012vf} to the $dS_4$ case with pseudo involution may shed more light on the unitarity in presence of fermions in $dS_4$ HS theory. An interesting option somewhat analogous to (but different from)  that discussed in \cite{Lang:2026inn} is that it may also require the  space-time doubling  by a parallel world.

 \section*{Acknowledgments}
We are  grateful to Chris Hull, Evgeny Ivanov, Yuihun Joung,  Yasha Neiman, Per Sundell, Valery Tolstoy,  and Francesco Toppan for stimulating discussions and Andrei Barvinsky for the correspondence. We thank Ofer Aharony, Theoretical High Energy Physics Group of Weizmann Institute of Science for hospitality at the final stage of this work. This work was supported by Theoretical Physics and
Mathematics Advancement Foundation “BASIS” Grant No 24-1-1-9-5.

\end{document}